\documentclass[preprint]{llncs}

\usepackage{listings}
\usepackage{color}
\usepackage[x11names]{xcolor}
\usepackage{graphicx, float}
\usepackage{tabularx}
\usepackage{colortbl}
\usepackage{booktabs} 
\usepackage{ifthen}
\usepackage{lmodern}
\usepackage[T1]{fontenc}
\usepackage{textcomp}
\usepackage{alltt, array, xspace, multicol} 
\usepackage{amsmath}
\usepackage{amssymb} 
\usepackage{stmaryrd}

\usepackage{tikz}
\usetikzlibrary{shapes,arrows,matrix}

\usepackage{hyperref}
\usepackage[capitalise]{cleveref} 
\hypersetup{
  colorlinks=true,%
  citecolor=red!80!black,%
  filecolor=red!80!black,%
  linkcolor=red!80!black,%
  urlcolor=blue!50!black
}

\renewcommand\cref[1]{\Cref{#1}}

\newboolean{showcomments}
\setboolean{showcomments}{false}
\ifthenelse{\boolean{showcomments}}
{ \newcommand{\mynote}[2]{\fbox{\bfseries\sffamily\scriptsize#1}
{\small$\blacktriangleright$\textsf{\emph{#2}}$\blacktriangleleft$}}}
{ \newcommand{\mynote}[2]{\!\!}}

\newcounter{mnotei} 
\newcommand{\mnote}[1]{%
{\scriptsize\textsf{\textcolor{blue}{$^{[\themnotei]}$}}}%
\marginpar{\scriptsize\textsf{\textcolor{blue}{n.\themnotei: #1}}}%
\stepcounter{mnotei} } 
\renewcommand{\mnote}[1]{} 

\usepackage{listings}
\newcommand*{\Comment}[1]{\hfill\makebox[6.0cm][l]{\rm\it \color{red!70!black}#1}}%
\newcommand{\ie}{\emph{i.e.,\xspace}}
\newcommand{\eg}{\emph{e.g.,\xspace}}

\newcommand{\pre}[1]{\texttt{#1}}

\definecolor{debuggray}{gray}{0.80}

\newcommand{\preciao}[3]{
\begin{figure}[b]
\fontsize{9}{10}
\fcolorbox{black}{white}{\begin{minipage}{\textwidth}
\begin{alltt}
#3%
\end{alltt}
\end{minipage}
}
\caption{#2}
\label{#1}
\end{figure}}

\newcommand{\ztag}[1]{\underline{\color{blue}{#1}}}

\begin{document}


\title{Reversible Language Extensions and their Application in Debugging}

\author{%
  Zo\'e Drey\inst{1} \and%
  Jos\'e F. Morales\inst{1} \and%
  Manuel V. Hermenegildo\inst{2,1}
}
\institute{%
  IMDEA Software Institute, Madrid (Spain)\and 
  School of Computer Science, T. U. Madrid (UPM), (Spain) 
}
\maketitle


\begin{abstract}
  A range of methodologies and techniques are available to guide the
  design and implementation of language extensions and domain-specific
  languages.
  A simple yet powerful technique is based on source-to-source
  transformations interleaved across the compilation passes of a base
  language. Despite being a successful approach, it has the main
  drawback that the input source code is lost in the process.
  When considering the whole workflow of program development (warning
  and error reporting, debugging, or even program analysis), program
  translations are no more powerful than a glorified macro language.
  In this paper, we propose an augmented approach to language
  extensions for Prolog, where symbolic annotations are included in
  the target program. These annotations allow selectively
  reversing the translated code.
  We illustrate the approach by showing that coupling it with minimal
  extensions to a generic Prolog debugger allows us to provide users
  with a familiar, source-level view during the debugging of programs
  which use a variety of language extensions, such as functional
  notation, DCGs, or CLP\{Q,R\}.
\end{abstract}




\section{Introduction}
\label{sec:introduction}

One of the key decisions when specifying a problem or writing a
program to solve it is choosing the right language. Even when using
recent high-level and multi-paradigm languages, the programmer often
still needs precise, domain-specific vocabulary, notations, and
abstractions which are usually not readily available.
These needs are the main motivation behind the development of
domain-specific languages, which enable domain experts to express
their solutions in terms of the most appropriate constructs.

However, designing a new language can be an intimidating task.  A
range of methodologies and tools have been developed over the years in
order to simplify this process, from compiler-compilers to visual
environments~\cite{mernik05}.
A simple, yet powerful technique for the implementation of
domain-specific languages is based on source-to-source
transformations. Although in this process the source and target
language can be completely different, it is frequent to be just
interested in some \emph{idiomatic extensions}, i.e., adding domain
specific features to a host language while preserving the availability
of most of the facilities of this language. Examples of such
extensions are adding functional notation to a language that does not
support it, adding a special notation for grammars (such as
Definite Clause Grammars (DCGs)~\cite{PereiraWarren80}), etc.
Such transformations have been proposed in the context of
object-oriented programming (\eg for Java,~\cite{nystrom:polyglot}), functional
programming (\eg for Haskell,~\cite{edsl}), or logic programming (the
\pre{term\_expansion} facility in most Prologs, or the extended
mechanisms
of~\cite{ciao-modules-cl2000,hermenegildo11:ciao-design-tplp})
In this approach, the language implementations provide a collection of
\emph{hooks} that allow the programmer to extend the compiler and
implement both syntactic and semantic variations.

An important practical aspect is that, in addition to appropriate
notation, the programmer also needs environments that help during
program development. In particular, basic tools such as editors,
analyzers, and, specially, debuggers are fundamental to productivity.
However, in contrast to the significant attention given to mechanisms
and tools for defining language extensions, comparatively few
approaches have been proposed for the efficient construction of such
development environments for domain-specific languages. In some cases
ad-hoc editors, debuggers, analyzers, etc.\ have been developed from
scratch.
However, this approach is time consuming, error prone, hard to
maintain, and usually not scalable to a variety of language
extensions.

A more attractive alternative, at least conceptually, is to reuse
the tools available for the target language, such as its debuggers or
analyzers. This can in principle save much implementation effort, 
in the same way in which the source-to-source approach leverages the
implementation of the target language to support the domain-specific
extensions.
However, the downside of this approach is that these tools will
obviously communicate with the programmer in terms of the target
language. Since a good part of the syntactic structure of the input
source code is typically lost in the transformation process, these
messages and debugger steps in terms of the target language are often
not easy to relate with the source level and then the target language
tools are not really useful for their intended purposes.
For example, a debugging trace may display auxiliary calls, temporary
variables, and obscure data encodings, with no trivial relation with
the control or data domain at the source level. Much of that
information is not only hard to read, but in most cases it should
be invisible to the programmer or domain expert, who
should not be forced to understand how the language at the source level
is embedded in the supporting language.

In this paper, we propose a method for recovering \emph{symbolically}
the source of particular translations (that is, \emph{reversing} them
and providing an \emph{unexpanded} view when required) in order to
make target language level development tools useful in the presence of
language extensions.  Our solution is presented in the context of
Ciao~\cite{hermenegildo11:ciao-design-tplp}, which uses a powerful
language extension mechanism for supporting several paradigms and
(sub-)languages.  We augment this extension mechanism with support for
symbolic annotations that enable the recovery of the source code
information at the target level. As an example application, we use
these annotations to parameterize the Ciao interactive debugger,
so that it displays domain-specific information, instead of plain
Prolog goals. Our approach requires only very small modifications in
the debugger and the compiler, which can still handle other language
extensions in the usual way.

The paper is organized as follows: \cref{sec:background} presents a
concrete extension mechanism  
and illustrates the limitations of the traditional translation
approach in our context.  \cref{sec:unexpansion} presents our approach
to unexpansion, and guidelines for instrumenting language extensions
so that the intervening translations can be reversed as needed into
their input source code. \cref{sec:application} presents the
application of the approach to the case of debuggers. Finally,
\cref{sec:relatedwork} presents related work and
\cref{sec:conclusions} concludes and suggests some future work.


\newcommand\sfun[1]{\llbracket \textit{#1} \rrbracket}
\newcommand\sfunx[2]{\llbracket \textit{#1} \rrbracket #2}

\section{Language extensions and their limitations}
\label{sec:background}

We present a concrete language extension mechanism based on
translations (the one implemented in the Ciao language) and then
illustrate the limitations of the traditional translation-based
extension approach in our context.  In
Ciao~\cite{hermenegildo11:ciao-design-tplp}, language extensions are
implemented through \emph{packages}~\cite{ciao-modules-cl2000}, which
encapsulate syntactic extensions for the input language, translation
rules for code generation to support new semantics, and the necessary
run-time code.  Packages are separated into compile-time and run-time
parts. The compile-time parts (termed \emph{compilation modules}) are
only invoked during compilation, and are not included in executables,
since they are not necessary during execution.
On the other hand, the run-time parts are only required for execution
and are consequently included in executables.
This phase distinction has a number of practical advantages, including
obviously the reduction of executable sizes.

More formally, let us assume that an extension for some language denoted
as $\mathcal{L}_e$ is defined by the package $\textit{PkgMod}_e$, and that
the compiler passes include calls to a generic expansion mechanism
$\sfun{expand}$, which takes a package, an input program in the source
language, and generates a program in the target language
$\mathcal{L}$.
That is, given $\sfunx{expand}{_e} =
\sfunx{expand}{(\textit{PkgMod}_e)}$, for a program $P_e \in
\mathcal{L}_e$ we can obtain the expanded version
$\sfunx{expand}{_e(P_e)} = P \in \mathcal{L}$. Note that in practice,
Ciao contains finely grained translation hooks, which allow a better
integration with the module system and the composition of
translations~\cite{composing-extensions-lopstr-11}. This level of
detail is not necessary for the scope of this paper, and thus, for the
sake of simplicity, the expansion will work on whole programs at a
time.

\newcommand\kw[1]{~\mbox{\tt #1}~}
\newcommand\rr[2]{\mbox{\textsf{\textbf{#1}}}\llbracket~#2~\rrbracket}
\newcommand\rrl[2]{\mbox{\textsf{\textbf{#1}}}\llbracket#2\rrbracket}

\vspace{1.5ex}
\noindent\textbf{Functional notation.}~~ We illustrate the translation
process in Ciao with an example from the \emph{functional notation}
package~\cite{functional-lazy-notation-flops2006}. This package extends
the language with \emph{functional}-like syntax for relations. Informally,
this extension allows including terms with predicate symbols as part of
data terms, while interpreting them as predicate calls \emph{with an
implicit last argument}. It also allows defining clauses in functional
style where the last argument is separated by a
\texttt{:=} symbol (as well as other functionalities, such as expanding
goals in the last argument after the body).  The translation can be
abstractly specified as a collection of rewrite rules such as:
\begin{displaymath}
  \begin{array}{crcl}
    \mbox{\ \ \ (Clauses) \ \ \ } & \rr{tr}{p(\bar{a}) \kw{:=} C \kw{:-} B} & = & (p'(\bar{v}, T) \kw{:-} \bar{v} \kw{=} \bar{a}, B, T\kw{=}C) \\
    \mbox{(Calls)} & \rr{tr}{q(\ldots p(\bar{a}) \ldots)} & = & (p'(\bar{a},T), q(\ldots T \ldots)) \\
  \end{array}
\end{displaymath}
The first rule describes the meaning of a clause in functional
notation, where $p'$ is the predicate in plain syntax 
corresponding to the definition of $p$ in functional notation (i.e.,
using \texttt{:=}). The second 
rule must be applied using a leftmost-innermost strategy for every $p$
function symbol that appears in the goal $q$, where $T$ is a new
variable (skipping higher-order terms).
If SLD resolution is used, the evaluation order corresponds to eager,
call-by-value evaluation (but lazy evaluation is possible and shown
in~\cite{functional-lazy-notation-flops2006}). 
We refer to the actual implementation later in this section.

\begin{figure}[t]
  \centering\small
  \begin{minipage}{0.45\linewidth}
    \emph{Source code (functional notation)}
\begin{lstlisting}[numbers=none,frame=r]
f(X) := X < 42 ?          
          (k(l(m(X))) * 3)
        | 1000.           
k(X) := X + 1.
l(X) := X - 2.
m(X) := X.

@\hspace{1ex}@
\end{lstlisting}
  \end{minipage}
  \hspace{1ex}
  \begin{minipage}{0.45\linewidth}
    \emph{Target code (plain Prolog)}
\begin{lstlisting}[numbers=none,frame=none]
f(X,Res) :- X < 42, !, 
    m(X, M), l(X, L), k(X, K),
    T is K * 3,
    T = Res.
f(X,1000).
k(X,Res) :- Res is X+1. 
l(X,Res) :- Res is X-2.
m(X,X).
\end{lstlisting}
  \end{minipage}
  \caption{Example translation for functional notation.}
  \label{fig:functional}
\end{figure}

\begin{example}
\label{ex:ex0}
  In ~\cref{fig:functional} we show an example program that defines a
  predicate \pre{f/2} in functional notation and its translation into
  plain Prolog code. Its body contains nested calls to \pre{k/2},
  \pre{l/2}, \pre{m/2}, and also syntactic sugar for a conditional
  (if-then-else) construct (using the syntax: \textit{CondGoal} \textrm{?}
    \textit{ThenExpr} \textrm{|} \emph{ElseExpr}). 
\end{example}

\subsubsection{Forgetful Translations and Loss of Symbolic Information.}
\label{sec:background:translation}

Both the standard compilation and the translations for language
extensions are 
typically focused on implementing some precise semantics during
execution. That is, the correctness of the translation guarantees
that for all programs $P_e \in \mathcal{L}_e$, the expected semantics
$\sfunx{exec}{_e}$ for that language can be described in terms of a
program $P \in \mathcal{L}$ and its corresponding execution mechanism
$\sfun{exec}$. That is, for all $P_e \in \mathcal{L}_e$ there
exists a $P = \sfunx{expand}{_e(P_e)}$ so that $\sfunx{exec}{_e(P_e)} =
\sfunx{exec}{(P)}$.

Most of the time, symbolic information at the source level is lost,
since it is not necessary at run time. In particular, such information
removal and loss of structure is necessary to perform
important program optimizations (\eg assigning some variables to
registers without needing to keep the symbolic name, its relation
to other variables in the same scope, etc.). When programs are not
necessarily executed, but manipulated at a symbolic level, the
translation-based approach is no longer valid on its own. For example,
assume a simple \emph{debugger} that interprets the source and allows
the user to inspect variable values at each program point
interactively. In this case the translation, as a program
transformation, must preserve not only the input/output
behaviour but also some other \emph{observable} features (such as line
numbers or variable names).

\begin{figure}[t]
     \centering
     \begin{tikzpicture}[
       code/.style={rectangle, draw}
       ]
  \matrix (m) [matrix of math nodes,row sep=3em,column sep=3em,minimum width=2em]
  {
    & \node [code] (pkgmod) {\textit{PkgMod}_e}; &
    & \node [code] (dbgmod) {\textit{DbgMod}};  & \\
    \node [code] (pe) {P_e}; &
    \node (comp) {\llbracket \textit{expand} \rrbracket}; &
    \node [code] (p) {P}; &
    \node (comp2) {\llbracket \textit{expand} \rrbracket}; &
    \node [code] (pdbg) {P_{dbg}}; \\
  };
  \path[-stealth]
    (pkgmod) edge (comp)
    (pe) edge (comp)
    (comp) edge (p)
    (p) edge (comp2)
    (comp2) edge (pdbg)
    (dbgmod) edge (comp2);
\end{tikzpicture}
\caption{The translation process and application of the standard debugger.}
\label{fig:debugprocess}
\end{figure}
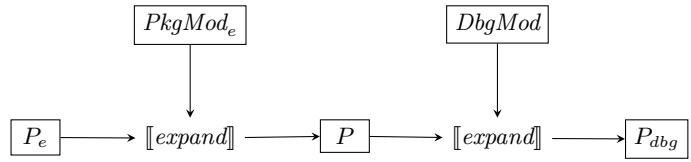

In order to explore the particular case of debuggers more closely,
\cref{fig:debugprocess} illustrates the translation process of a
source program, using a compilation module $\textit{PkgMod}_e$
containing the translation rules for extension $e$. If the developer
asks the Ciao interpreter to debug this program, further
instrumentation is applied that is also defined in part as a language
extension, $\textit{DbgMod}$ in \cref{fig:debugprocess}; this
instrumentation customizes the code by encapsulating it into a
predicate that specifies whether a part of the code is \emph{spy-able}
or not. The following example illustrates in a concrete case 
the limitations of this process.

\preciao{fig:ex0debug}{Excerpt of the display of the interactive debugger.}{
   2  2    Call: f(3,\_6378) ? \\ 
   3  3    Call: <(3,42) ? \\ 
   4  3    Call: m(3,\_6658) ? \\ 
   5  3    Call: l(3,\_6663) ? \\ 
   6  4    Call: is(\_6663,3-2) ? \\ 
...\\
   9  3    Call: is(\_6673,2*3) ? \\ 
  10  3    Call: =(\_6378,6) ?  
}

\begin{example}[Interactive debugging]
  Consider the code and transformation of~\cref{ex:ex0}. If the
  target-level debugger is used without any other provision, following
  the process of \cref{fig:debugprocess}, debugging a call to
  \pre{f(3,T)} amounts to debugging its translation, as illustrated in
  the trace of~\cref{fig:ex0debug} (the exit calls are omitted in
  order to save space).  The problem of this trace is twofold: first,
  the interactive debugging does not make explicit the actual source-level
  predicate that is currently being tested. Second,
  understanding the trace forces the developer to make the mental
  effort of analyzing the debugged data and mapping it back to the
  source code. This effort increases if the source code contains
  operators that do not exist on the target (Prolog) side. The first
  case can be easily overcome when operator definitions are shared,
  \eg using a graphical editor and catching the operator with the line
  number and the occurence number of the call. However, the second
  case implies remembering the mapping between the source and the
  target operator. Furthermore, things get even more tedious and
  intricate when one instruction in the source language is
  translated into a composition of goals.
\end{example}
 

\section{Building reversible extensions}
\label{sec:unexpansion}

In this section we provide an informal definition of \emph{unexpansion} with
respect to a language extension. We then present guidelines in order
to instrument a compilation module for such a language extension. The
purpose of this instrumentation is to drive the process of
reconstructing a program in terms of the language extension (or
\emph{source} language) in which the program is written. Through this
mechanism, a language extension can be made \emph{reversible}.  To
illustrate our objective, we apply the guidelines and parameterize one
of the translation rules used in the functional notation extension.

\subsection{A correspondence between expansion, unexpansion, and observers}

We use the term \emph{unexpansion} to designate the inverse of the
expansion $\sfun{expand}_e$, that is, the recovering of the original
$P_e$ source program
from $P$. Unfortunately, this inverse is rarely a one-to-one mapping.
For example, \verb|f(3,T)| in $\mathcal{L}$ corresponds to both
\verb|T=f(3)| and \verb|f(3,T)| (with \verb|f/1| using functional
notation). For another example, a clause can either be translated
in one or many clauses, as depicted in Figure~\ref{fig:functional} 
for  \verb|f| in functional notation.

Not existing a unique solution can be confusing for the
user and impractical for automatic transformations.
However, the most important use of unexpansion in our context is to
observe the behavior of only certain program aspects at the source
language level. In this case, unexpansion seems more treatable. For
that purpose we define the term \emph{observer} accordingly:
an \emph{observer} is an interface that provides some specific
source-level information about a particular program. The observer can
be either static or dynamic. Specifically, we can consider as
observers monitors (\eg interactive debuggers, tracers, and profilers)
for dynamic observation, and verifiers (\eg static analyzers and model
checkers) for static observation. Thus, a source-level view may
correspond to the current instruction being invoked in an interactive
debugger, or to a trace of the memory state, in a tracer, or perhaps
the dependencies between the program variables, in a static analyzer,
all of them represented in terms of the source language abstractions.

\begin{figure}[t]
    \centering
    \begin{minipage}{0.45\linewidth}
      \begin{tikzpicture}[
        code/.style={rectangle, draw}
        ]
  \matrix (m) [matrix of math nodes,row sep=7em,column sep=9em,minimum width=2em]
  {
     \node [code] {P_e}; & \node [code] {P}; \\
     \node [code] {V_e}; & \node [code] {V}; \\};
  \path[-stealth]
    (m-1-1) edge node [left] {$\textsf{Obs}_e(i)$} (m-2-1)
            edge node [below] {$\sfun{expand}_e$} (m-1-2)
    (m-1-2) edge node [left] {$\textsf{Obs}(i)$} (m-2-2);
\end{tikzpicture}
    \end{minipage}
    \begin{minipage}{0.45\linewidth}
      \begin{tikzpicture}[
        code/.style={rectangle, draw}
        ]
  \matrix (m) [matrix of math nodes,row sep=7em,column sep=9em,minimum width=2em]
  {
     \node [code] {P_e}; & \node [code] {(P, \textit{Sym})}; \\
     \node [code] {V_e}; & \node [code] {V}; \\};
  \path[-stealth]
    (m-1-1) edge node [left] {$\textsf{Obs}_e(i)$} (m-2-1)
            edge node [below] {$\sfun{expand}^{\textit{sym}}_e$} (m-1-2)
    (m-1-2) edge node [right] {$\textsf{Obs}^{\textit{sym}}(i)$} (m-2-2)
    (m-2-2) edge node [above] {$?$} (m-2-1.east|-m-2-2)
    (m-1-2) edge [dashed] node [right] {~~$\textsf{Obs}^{\textit{sym}}_e(i)$} (m-2-1);
\end{tikzpicture}
    \end{minipage}
    \caption{Observation problem at the source level (left);
      Observation using symbolic information (right).}
    \label{fig:unexpansion}
\end{figure}
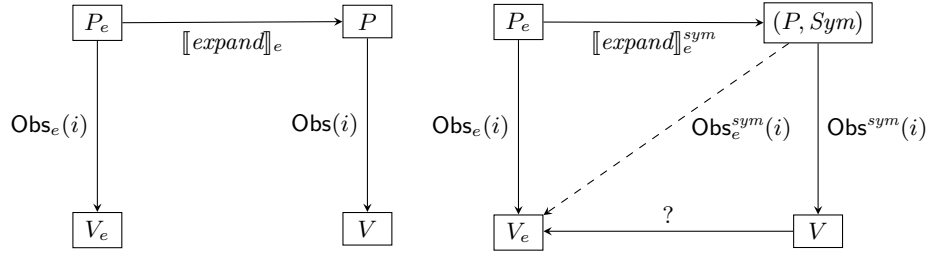 

The correspondance between expansion and unexpansion, in the context
of an observer, is sketched in Figure~\ref{fig:unexpansion}. We assume
that we have observers $\textsf{Obs}_e(i)$ and $\textsf{Obs}(i)$ for
the source and target languages, respectively. We denote by $i$ some
particular observable aspect and by $V$ the aspect (\eg ``line
numbers'' and an integer). On the left diagram we depict the
impossibility of getting information at the $\mathcal{L}_e$ level in
general. 
To provide the programmer with source-level observers, our approach
relies on extending the expansion ($\sfun{expand}^{\textit{sym}}_e$)
with additional symbolic information (which can be significantly 
smaller than the sources). Then, observers
$\textsf{Obs}^{\textit{sym}}(i)$ can retrieve $V$ (\eg a single
number encoding the row and columns) and map it back to $V_e$ (\eg
the row and columns). This composition provides an effective
$\textsf{Obs}^{\textit{sym}}_e(i)$.

We now propose guidelines for easily instrumenting the translation
module of a language extension, in such a way that observers can be
parameterized with respect to this instrumentation.

\subsection{Instrumentation of a compilation module}

Instrumenting a compilation module involves annotating its translation rules
with source code information that can then be used by an observer (\ie the
debugger in our application example). 
We illustrate the instrumentation process on the functional extension
example.

\subsubsection{Guidelines.}

The first step in making a language extension reversible is to determine which
parts of the source code need to be kept available in the expansion process.
The second step is to determine how and where to propagate this information,
so that it can be  accessed whenever the developer requires observation during 
program execution. The third step is to determine the representation of the 
observable data.

\paragraph{Event and data analysis.} What events do we want to
observe? What do we want to observe about them? These selections
should be useful for following the control flow and state changes
during program execution. For example, in a $\lambda$-calculus-like
language, the definition and the application of a function are two of
the key elements to follow in order to debug a
program~\cite{tolmach}. As another example, in a goal involving
expressions in functional notation, the debugger must be aware of
which positions correspond to data terms and which positions to
predicate calls.

\paragraph{Decomposition.} How is a source statement decomposed into
target code? The answer to this question implies in part how the data
that we want to observe should be propagated. For example, while the
generic debugger may step through a number of target-level statements,
a source-specific debugger may have to consider a single source
statement as corresponding to all those steps. This applies for
example in the conditional statement \pre{C ? A | B} of the functional
notation, where \pre{A} is translated into an (at least) two-goal
target code segment.

\paragraph{Representation.} How should the data to be observed be
represented?  In a purely syntactic extension, data always represents
elements of the concrete syntax. Nevertheless, it is interesting to
consider this question when displaying the runtime context, such as
the state of the memory, for semantic extensions.

For example, in a CLP\{Q,R\} extension, variables are bound at
run-time to complex terms attached to attributed variables which
reflect the internal, low-level representation of the constraint
store, while what the programmer would like to see is a symbolic
representation of the constraints among the variables in the source
constraint language.

\subsubsection{Instrumentation in action.}

To instrument the translation rules we propose to annotate the target
parameter of each rule (i.e., the argument in which the code generated
by the translation is returned). This annotation (which we call the
\emph{meta-annotation}) is defined as a macro which provides the
symbolic information to drive the process of recovering source code
data within the observer. It may contain any data written in a prolog syntax,
enabling to recover some source level information. 

For example, such annotation could be 
a list of variables and a function enabling to recover their value
in the source level notation from the target context (its environment
and store), or a single string to be displayed at the 
observer's output at run time.

We currently distinguish two types of
meta-annotations: the \pre{\$clause\_info} annotation, which is
wrapped around target clauses, and the \pre{\$goal\_info}
meta-annotation, which is wrapped around target goals. The purpose of
each of these meta-annotations is to gather symbolic information to
recover a source-level statement or a source-level call, respectively.
Additionally, this distinction enables to handle clauses and goals
properly, in particular to retrieve their location in source modules.

A meta-annotation takes two arguments: the first argument is the
wrapped element (i.e., the original clause or goal(s) generated by the
transformation), and the second one provides symbolic information
enabling to recover an ``observable'' representation of the wrapped
element, according to what the extension designer wants the programmer
to observe. 
We illustrate this annotation process with Example~\ref{ex:instr}. 

\begin{example}
\label{ex:instr}
Let us consider the translation rule for clause declarations in the
functional notation package. This rule, named \pre{defunc}, translates
such clause declarations into a set of clauses:
\begin{lstlisting}[frame=none,escapechar=@,numbers=none]
defunc((FuncHead := FuncValOpts), Clauses) :- 
    FuncValOpts = (FuncVal1 | FuncValR), !, 
    Clauses = [Clause1 | ClauseR],
    defunc((FuncHead := FuncVal1), Clause1),     @\Comment{(1)}@
    defunc((FuncHead := FuncValR), ClauseR).     @\Comment{(2)}@
\end{lstlisting}
\end{example}

The \pre{FuncHead} part on the left corresponds to a predicate
declaration; the \pre{FuncValOpts} part on the right corresponds to
goal invocations (this results from the data analysis guideline).
Notice that the declaration is decomposed into many goals (marked
\emph{(1)} and \emph{(2)}) if the \pre{|} operator appears inside its
right part. Therefore, the translation needs to be adapted slightly,
in order to indicate to the debugger that the declaration is to be
treated as a single one. As illustrated in Example~\ref{ex:syminstr}
below, the resulting adaptation amounts to creating an intermediate
predicate (\pre{defunc\_rec}, not really necessary in this simple
case), and to annotating the \pre{defunc} rule (this results from the
decomposition guideline).  Note that the \pre{\$clause\_info} wrapper
effectively groups all the clauses into which the definition is
expanded, and this can be detected by the observer which will then
treat it as a single clause.

The symbolic information attached to the annotation is represented by
the contents of variable \pre{SI}. This variable is handled by an
observer, according to the nature of the program view it aims to
provide. For example, line numbers, variables or function names can be
attached to it.  It can even be left as a free variable, in cases
where the observer can automatically retrieve the information.

This approach based on meta-information enables us to envision a range of
program views, from simple syntax recovery to high-level representation of
analysis results: annotations can be enriched with source-specific procedures
to handle various representations of the target program, enabling different
instantiations of the meta-annotation variable. They can even hold procedures
that perform advanced computations parameterized with the symbolic information (\eg
counting the number of times a function is invoked).

\begin{example} The instrumentation of the translation rule for declarations in
functional notation writes as follows:
\label{ex:syminstr}

\begin{lstlisting}[numbers=none,frame=none, mathescape=no]
defunc((FuncHead := FuncValOpts), $clause_info(Clauses, SI)) :-
    defunc_rec((FuncHead := FuncValOpts), Clauses),
    SI = (FuncHead := FuncValOpts).

defunc_rec((FuncHead := FuncValOpts), Clauses) :- 
	    FuncValOpts = (FuncVal1 | FuncValR), !,
    Clauses = [Clause1 | ClauseR],
    defunc_rec((FuncHead := FuncVal1), Clause1),
    defunc_rec((FuncHead := FuncValR), ClauseR).
\end{lstlisting}
\end{example}

The same instrumentation method applies to goals, as outlined in the
schema of Figure~\ref{fig:tree}, which depicts a declaration of the
form \texttt{f(X) := }\textit{Cond}\texttt{ ? }$B_1$\texttt{ |
}$B_2$. In this figure, the variable names \pre{S$x$} correspond to
symbolic information for some program elements (like goals or
clauses), and the expressions $\rrl{tr}{x}$ correspond to a
translation of the term $x$. To avoid the overloading of the
compilation module with annotations, symbolic information can be
stored in a specific table.

\newcommand{\ikw}[1]{{\color{red!70!black}#1}} 
\newcommand{\iv}[1]{{\color{DodgerBlue3}#1}} 
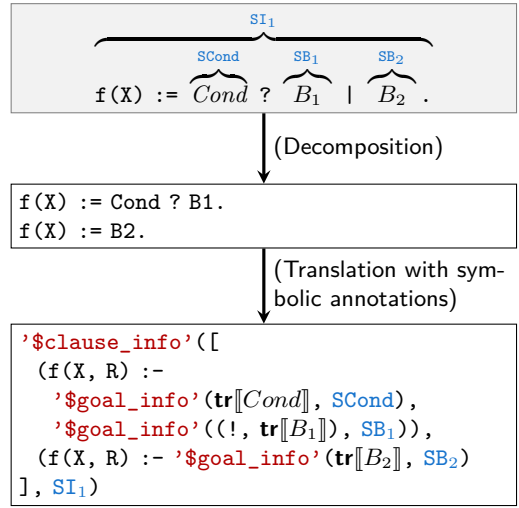
\begin{figure}[t]
    \centering
    \begin{tikzpicture}[
      node distance=2.1cm,
        code/.style={rectangle, draw, text width=20em, align=center}
        ]
        \node [code, color=black!50, fill=black!5] (c1) {
          $\overbrace{
            \texttt{f(X) := }
            \overbrace{\textit{Cond}}^{\texttt{\iv{SCond}}}
            \texttt{ ? }\overbrace{B_1}^{\texttt{\iv{SB\ensuremath{_1}}}}
            \texttt{ | }\overbrace{B_2}^{\texttt{\iv{SB\ensuremath{_2}}}}
            \texttt{.}
          }^{\texttt{\iv{SI\ensuremath{_1}}}}$
        };
        \node [code, below of=c1, align=left] (c2) {
          \texttt{f(X) := Cond ? B1.}\\
          \texttt{f(X) := B2.}
        };
        \node [code, below of=c2, align=left, yshift=-4ex] (c3) {
\texttt{\ikw{'\$clause\_info'}([}\\
~~\texttt{(f(X, R) :-}\\
~~~~\texttt{\ikw{'\$goal\_info'}($\rrl{tr}{Cond}$, \iv{SCond}),}\\
~~~~\texttt{\ikw{'\$goal\_info'}((!, $\rrl{tr}{B_1}$), \iv{SB$_1$})),}\\
~~\texttt{(f(X, R) :- \ikw{'\$goal\_info'}($\rrl{tr}{B_2}$, \iv{SB$_2$})}\\
\texttt{], \iv{SI$_1$})} 
        };
        \path [-stealth, very thick] (c1) edge node[right] {\sf (Decomposition)} (c2) ;
        \path [-stealth, very thick] (c2) edge node[right, text width=10em] {\sf (Translation with symbolic annotations)}  (c3);
      \end{tikzpicture}
    \caption{Instrumented translation of a clause in functional notation.}
    \label{fig:tree}
\end{figure}


\section{Application to the interactive debugger}
\label{sec:application}

We now illustrate the use of a reversible language extension to
parameterize the generic interactive debugger of Ciao. We describe the
modifications performed on the compiler and on the debugger, and show
the resulting source-level trace for our initial example of
Figure~\ref{fig:functional}.

\subsection{Implementation details}

\begin{figure}[t!]
    \centering
    \includegraphics[width=\columnwidth]{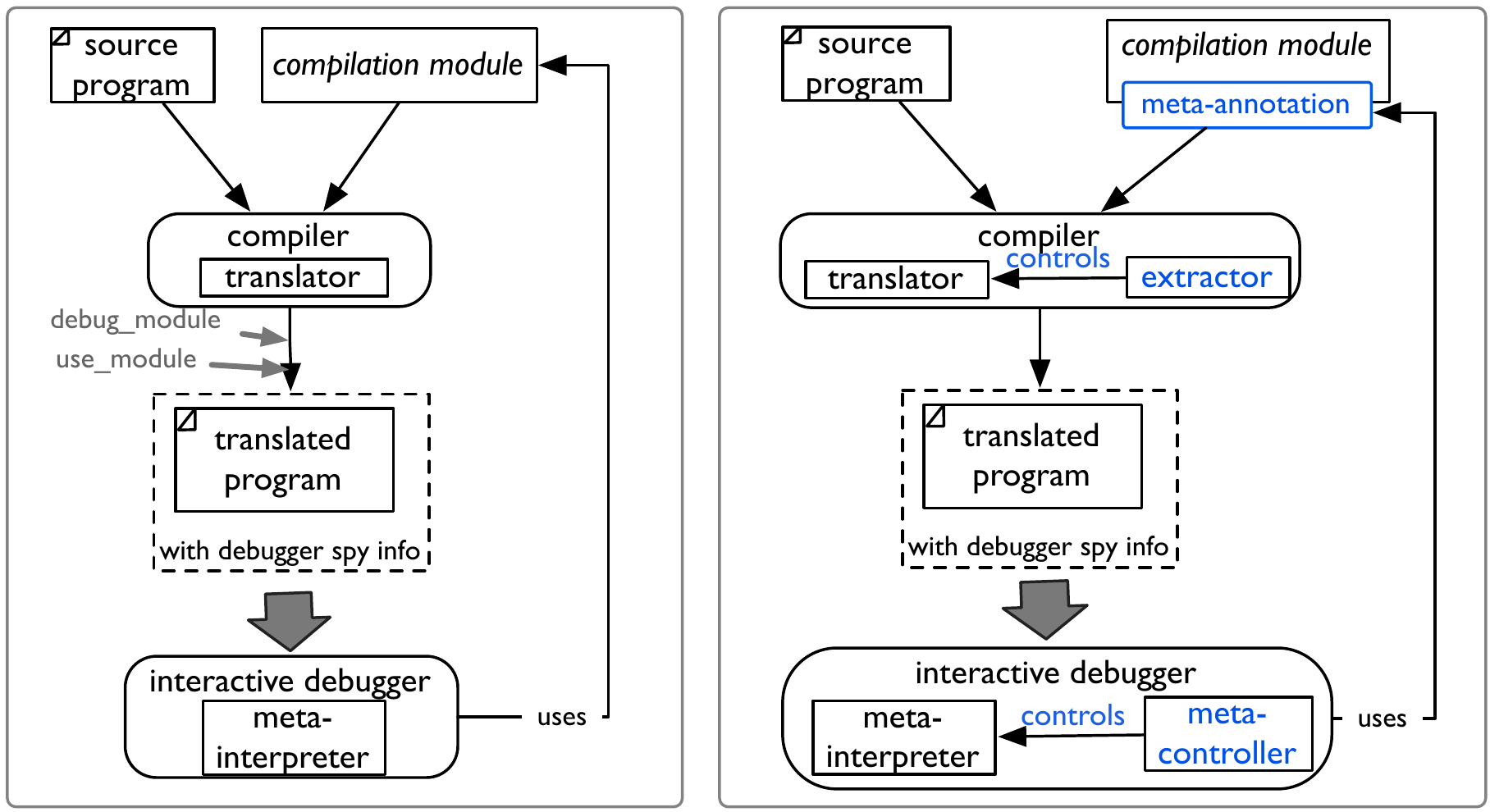}
    \caption{Implementation: original (left) vs.\ customized (right)
    infrastructure.}
    \label{fig:double-process}
\end{figure}

The overall process of making program behavior observable at the
source level through a debugger and reversible expansion is depicted
in Figure~\ref{fig:double-process}.

The compiler is responsible for applying both the debugger compilation
module and the source language compilation module. Prior to applying
the translation rules, it extracts the elements corresponding to
sentences, clauses, and goals. During this step, information to locate
the source program instructions are saved, such as the module name,
the line numbers for sentences, and the name of the goal being called.
Then, sentences, clauses, and goals are translated according to the
specifications of the corresponding compilation module. To enable the
handling of the \pre{\emph{term}\_info} meta-annotations in Ciao, the
translation step (represented by the \pre{translator} box in
Figure~\ref{fig:double-process}) of the compiler needs to be
customized. This is done by performing an extraction step (represented
by the \pre{extractor} box in Figure~\ref{fig:double-process}, right
part) that modifies the translation process when a meta-annotation is
encountered.

In the case of the debugger, the required symbolic information
corresponds to a source node (\eg \pre{k(X) := X + 1} as in
Figure~\ref{fig:functional}). As a result, the extraction process
consists solely of storing each source node (either a clause or a
goal) before its expansion.

Once the source-level information is extracted and mapped to the
appropriate target term (or composition of target terms, \emph{cf.}
the guidelines in Section~\ref{sec:unexpansion}), it is interpreted by
the debugger. To step through the source code instead of the target
code, the debugger is equipped with a \pre{meta-controller}, which
checks the presence of a meta-information call at the level 
of the translated program, and displays a trace
step accordingly. In particular, it is responsible for locating the
name of the target goal in the source nodes corresponding to this
goal. Since the compiler provides the source code information as a
Prolog term, this localization is straightforward. When a goal invoked
in the debugger has not been annotated (with
\pre{\$goal\_info}), the meta-controller looks into the last \pre{\$clause\_info}
meta-annotation, and looks for the name of this goal inside this
meta-annotation. Otherwise, the standard, expanded debug information
is displayed.
 
\subsection{Source-level tracing: the functional example revisited}

\preciao{fig:ex0newdebug}{An excerpt of the debugger trace, customized with source information.}{
   2  2    Call: ex0:f(3,\_6371) ? \\ 
   3  3    Call: f(3) := 3 \ztag{<} 42 ? k(l(\ztag{m}(3)))*3 | 1000 ? \\ 
   4  4    Call: f(3) := 3 < 42 ? k(l(\ztag{m}(3)))*3 | 1000 ? \\ 
   5  5    Call: m(3) := 3 ? \\ 
   6  4    Call: f(3) := 3 < 42 ? (k(\ztag{l}(m(3)))*3 | 1000 ? \\ 
   7  5    Call: l(3) := 3 \ztag{-} 2 ? \\ 
   8  4    Call: f(3) := 3 < 42 ? \ztag{k}(l(m(3)))*3 | 1000 ? \\ 
   9  5    Call: k(1) := 1 \ztag{+} 1 ? \\ 
   10  3    Call: f(3):= 3 < 42 ? k(l(m(3))) {*} 3 | 1000 ? \\
   2  2    Exit: ex0:f(3,12) ? 
}

With this instrumentation, Example~\ref{ex:ex0} is now debugged in
source code terms, as illustrated in
Figure~\ref{fig:ex0newdebug}. Note that the debugger now displays the
complete declaration (see second line) defining \pre{f}, instead of a
single part of a clause (see the second line in
Example~\ref{ex:ex0}). When a function evaluation returns a value
(which is the case of all the functions \pre{f/1}, \pre{k/1},
\pre{l/1}, \pre{m/1}), intermediate unifications are performed by the
generic debugger.  When the debugger is instrumented with a meta
controller (\ie the handler of meta-annotations), these unification
steps are ignored (skipped over), since they have no representation in
the original source code.


\section{Related Work}
\label{sec:relatedwork}

There exist frameworks and generative approaches that facilitate the
development of DSL tools for programming, including
debuggers~\cite{eclipse,tide}. For example, the Eclipse Integrated
Development Environment~\cite{eclipse}, provides an API and an
underlying framework that can greatly help in the development of a
debugger~\cite{eclipsehowtodebug}.  Emacs 
is another example of such environments, with facilities in the same
line as
Eclipse.  However, 
these tools are large and have a significant learning curve, and, more
importantly, their facilities are centered more around the graphical
navigation of the source code and interfacing with a command-line
debugger, while the focus of our work is on bridging syntactic or
semantic aspects between two sides of a translation, within such a
command-line debugger.
In that sense our work is complementary to (and in practice combines
well with) the facilities in Eclipse, Emacs, and related environments.
Generative approaches have been suggested (\eg based on aspect weaving
into the language grammar~\cite{debugaspect05}) in order to reduce
developer burden when using intricate APIs.  

However, none of these approaches provide a methodology for developing
reliable and maintainable debuggers. As a result, the development of
debuggers has remained difficult, inciting DSL tool developers to
implement ad-hoc solutions,
through extension-specific modifications and adaptations of the
debugger code. For example, SWI-Prolog includes a graphical debugger %
for Prolog with built-in support for DCGs and Logtalk
programs~\cite{swipl-debug}. As mentioned in the introduction, this
approach results in useful debuggers but which are specific to
concrete extensions. As a result, they have to be modified again for
other transformations.

Our objective has been to develop a more general approach, which we
have illustrated by applying the same 
methodology to several extensions including functional notation, DCGs, 
and CLP\{Q,R\}.

Lindeman \emph{et al.}~\cite{lindeman11} have proposed recently a
declarative approach to defining debuggers. To this end, they use
SDF~\cite{sdfasf}, a rewriting system, to instrument the abstract
syntax tree with debugging annotations. However, it does not seem
obvious that their approach could be applied to other observer
tools. Indeed, instrumentation is achieved by providing
debugger-specific information, in the form of events. In contrast, our
instrumentation process makes it possible to easily add and handle
different kinds of meta-information.

Unexpansion and decompilation only differ in the hypothesis used in
decompilation: that the original source code may not be available. It
is interesting however to compare to existing related decompilation
approaches. Bowen~\cite{decompilationprolog} proposes a compilation
process from Prolog to object code which makes it possible to define
decompilation as an inverse call to compilation, provided some
reordering of calls is performed. Gomez \emph{et
  al.}~\cite{mod-decomp-jist09} also propose a decompilation process for
Java based on partial evaluation. However, these approaches have not
been designed to be applicable to a large class of different language
extensions.  More generally, while it is in theory possible (although
predictably hard with current technology) to implement fully
reversible transformations, this approach runs into the problem that
such inversions are non-deterministic in general, in the sense that a
given target code can be generated from multiple source texts.
Presenting the programmer with a different code that what is in the
source program could be even more confusing that debugging the target
code directly.

More similar to our solution is the approach of Tratt~\cite{tratt08},
which also targets language extensions, and where source information
is injected into the abstract syntax tree of the source program. This
information is exploited to report errors in terms of the language
extension. However, they only discuss how to inject such information
in the syntax tree, and do not explain how to use this information
when building or adapting tools.

The macro-expansion passing style~\cite{dybvigEPS} approach makes it
possible to easily implement observers. Our approach differs from this
one in the reliance on the existing generic debugger (Ciao's in our
examples), and concentrates instead on what changes are required in
the debugger and the extension framework in order to handle
meta-information for unexpansion in a way that is independent from
the concrete language extension.

As a conclusion, we believe that our process proposal could be
extended to other Prologs, as the meta-annotations enable to
hold symbolic information that is made available in most Prolog compilers,
e.g., line numbers or variable names.


\section{Conclusion and future work}
\label{sec:conclusions}

We have presented a generic approach that enables a debugger for a
target language to display trace information in terms of the language
extension in which a source program is written, using the Ciao
debugger as an example. The proposed approach is based on an extension
of the usual mechanisms for term expansion, and in particular of their
modular implementation in Ciao through \emph{packages}. Specifically,
we define a methodology for making relevant parts of the source text
and other characteristics at the target level by enriching the
translation rules.  We have shown that the compiler and the debugger
require only small adaptations in order to take this mechanism into
account and that these adaptations are generic in the sense that while
the transformation rules are of course specific to the extension, the
compiler and debugger themselves do not require further modification,
for what is arguably a usefully large class of extensions. In
particular, in the paper we have illustrated this approach by applying
it on the functional notation. In the system,
we have successfully applied it also to the DCG and CLP\{Q,R\}
constraint packages.

In future work, we plan to extend the flexibility of the approach by
enriching the annotations, and being able to provide different
annotations for different purposes. 
Also, we feel that this initial work on augmenting the language
extension mechanism already provides us gives with the basis for
adapting the Ciao pre-processor so that for example errors, warnings,
and other reports are made in terms of the source, domain-specific
language, for different extensions, without requiring further
modification of the pre-processor itself.  The same would apply of
course to the auto-documenter. 

Finally, we could leverage Kishon \emph{et al.}'s
framework~\cite{kishonmonitoring} to check the soundness
of our approach with regard to the intended semantics of a language extension.
Doing so would also enable to show the equivalence
between the behavior of an ad-hoc source level debugger and our customization
of the target level debugger.

            
\subsection*{Acknowledgments}

The research leading to these results has received funding from the
Madrid Regional Government under CM project P2009/TIC/1465
(PROMETIDOS), and the Spanish Ministry of Economy and Competitiveness
under project TIN-2008-05624 {\em DOVES}.


\bibliographystyle{abbrv}

\bibliography{dslysis-final}
\end{document}